\documentclass[fleqn,preprint,showpacs,preprintnumbers,amsmath,amssymb]{revtex4}
\usepackage{graphicx}
\usepackage{dcolumn}
\usepackage{bm}

\begin{document}

\title{Comment on "Interaction of two solitary waves in quantum electron-positron-ion plasma" [Phys. Plasmas \textbf{18}, 052301 (2011)]}
\author{M. Akbari-Moghanjoughi}
\affiliation{ Azarbaijan University of
Tarbiat Moallem, Faculty of Sciences,
Department of Physics, 51745-406, Tabriz, Iran}

\date{\today}
\begin{abstract}
Recently, Yan-Xia Xu, et al. in the article Ref. [Phys. Plasmas \textbf{18}, 052301 (2011)] have studied the effects of various plasma parameters on interaction of two ion-acoustic solitary waves in an unmagnetized three-dimensional electron-positron-ion quantum plasma. They have used the extended reductive perturbation technique, the so-called, extended Poincare´-Lighthill-Kuo (PLK) technique, to deduce from the model governing the quantum hydrodynamics (QHD) differential equations leading to the soliton dynamical properties, namely, Korteweg-de Vries evolution equations (one for each wave) and coupled differential equations describing the phase-shift in trajectories of solitons due to the two dimensional collision. The variation of the calculated collision phase-shifts are then numerically inspected in terms of numerous plasma fractional parameters. In this comment we give some notes specific to the validity of the results of above-mentioned article and refer to important misconceptions about the use of the Fermi-temperature in quantum plasmas, appearing in this article and many other recently published ones.
\end{abstract}


\maketitle

Recently a major research focus appears on investigation of nonlinear wave propagations in ultra-cold dense plasmas, so-called the quantum plasmas, due to their applications in both astrophysical and laboratory \cite{Chandra, Manfredi, Gardner} environments. Many of the investigations using the quantum hydrodynamics (QHD) model rely on the assumption of complete degeneracy or the zero-temperature Fermi-gas model with a well defined Fermi energy for the corresponding degenerated plasma species. Shukla and Eliasson \cite{Shukla} have given an extensive review on different aspects of a quantum plasma. Conveniently, the Fermi-energy is defined through a somehow artificial quantity called the Fermi-temperature ($E_F=k_B T_F$) \cite{landau} which of course is not a real physical quantity. Unfortunately, however, we see that this quantity (or the corresponding fractional parameters) inaccurately appears in the numerical evaluations (or even used to compared to plasma temperatures, for example, $T_{Fe}/T_i$) in many articles without noticing the fact that this quantity has nothing to do with the physical plasma temperature. It is becoming widespread (see the references mentioned below) to evaluate the fractional Fermi-temperatures independently in QHD plasma model. As it is clearly obvious the Fermi-temperature is just a symbol used to label the highest possible energy of a particle (Fermi-energy) in a quantum plasma and is related to the number density of the corresponding quantum plasma specie, $n$ \cite{Akbari}. In the recently published article \cite{Yan}, we find that the fractional parameters, $\sigma$ (positron to electron Fermi-temperature ratio) and $\mu$ (positron to electron number-density ratio) should not be treated as independent plasma parameters, but we have, $\mu=1/(1-\sigma^{2/3})$ instead. It has been shown \cite{Akbari} that, such inaccuracy can lead to completely converse results.

To this end, lets consider the article by E. F. El-Shamy et al. \cite{Elshamy} where they have studied the head-on collision of ion-acoustic waves in an unmagnetized electron-positron-ion plasma in the framework of one-dimensional Thomas-Fermi approximation for electrons and positrons and they consider hot isothermal or adiabatic ions. The normalization in their work clearly indicates a well defined Fermi-energy, $E_F$, which implies that the plasma is in a fully degenerate state with a very small electron/positron thermal temperature (compared to the corresponding Fermi-temperatures), i.e. $T_{thj}\ll T_{Fj}$ ($j=e,p$), where $T_{Fj}$ is the $j$-specie Fermi-temperature which is also well defined through the relation $T_{Fj}=E_{Fj}/k_{B}$. As mentioned above, in the fully degenerate configuration, the Fermi-energy and consequently the Fermi-temperature of each degenerate plasma specie (electron/positron) is fixed via the number-density of that specie in the system, through the following relation
\begin{equation}\label{T}
{T_{Fj}} = \frac{E_{Fj}}{k_B} = \frac{{{\hbar ^2}}}{{2{m_j}{k_B}}}{(3{\pi^2}{n_{j0}})^{2/d}},
\end{equation}
where, $d$ is the dimensionality. Using the notations given in Ref. \cite{Elshamy} for positron to electron equilibrium number-density ratio, $\alpha=n_{p0}/n_{e0}$, and electron to positron Fermi-temperature ratio, $\sigma=T_{Fe}/T_{Fp}$, it is easily concluded from Eq. (\ref{T}) that the fractional plasma parameters $\alpha$ and $\sigma$ are not independent, but, they are related in the form $\sigma  = {\alpha^{- 2/d}}$ as neglected in Ref. \cite{Elshamy}. The preceding argument indicates that one can not fix $\sigma$ or $\alpha$ and vary independently the other, as it is employed in Figs. 1, 2 and 3 of Ref. \cite{Elshamy}. Also, for example, the value of $\alpha=0.2$ given in Fig. 4 of Ref. \cite{Elshamy} should correspond to the value $\sigma= 25$ and not the value $\sigma=1$.

Having in mind that $\sigma  = {\alpha^{- 2}}$ in the one-dimensional case and borrowing the coefficients $A$, $B$, $C$ and $D$ from Ref. \cite{Elshamy}, the variations of collision phase-shift with respect to the fractional positron to electron density, $\alpha$, and the ion-temperature to electron Fermi-temperature ratio, $\sigma_F$ (with identical values and plot ranges used in Ref. \cite{Elshamy}) is shown in Fig. 1. The plots with labels $a$, $b$, $c$ and $d$ in Fig. 1, here, correspond to the Figs. $1$, $2$, $3$ and $4$ in Ref. \cite{Elshamy}, respectively. The solid curves in Fig. 1 (shown here) corresponds to the adiabatic and the dashed curves refer to the isothermal case.

As it is observed from Fig. 1(a)-1(c), the collision phase-shift increases with increase in the relative positron density, $\alpha$, when the fractional ion-temperature, $\sigma_F$, is fixed. It is also remarked from Fig. 1(a) that, the values of the phase-shift for a given value of $\alpha$ is comparably lower in adiabatic-ion case relative to that of isothermal one. These result are quite contrary to the ones given in Figs. 1, 2 and 3 of Ref. \cite{Elshamy}. Also, the variation of phase-shift with respect to $\sigma_F$ in the cases of isothermal and adiabatic ions for a fixed relative positron density, $\alpha$, is completely converse to the Fig. 4 in Ref. \cite{Elshamy}.

Unfortunately, the same inaccuracy applies to numerous articles, \cite{Khan1, Khan2, Ali, Moslem, Mushtaq1, Mushtaq2, Masood, Abdelsalam, Yue, Saeed, Zhenya, Akhtar}.

Concerning the article Ref. \cite{Yan} the similar problem arises and the authors should revise all their plots correspondingly, although, even the given plots can not be produced with the parameters given in article ($A$, $B$, $C$, $D$, etc. parameters in the article). Another problem concerning this article is that the values given in the caption of Figs. 1 and 2 for instance are not consistent with Eq. (A25). Equation (A25) of the article gives imaginary values for speed, $c_1$, for some range of $\mu$ and $\sigma$ values. This may be due to ignoring of the $\sigma$ and $\mu$ relation. It is further noted that, the $A$, $C$, $D$ coefficients are not invariant under the interchange of the two solitons as it should be for a symmetric collision.

\newpage

\begin{figure}[ptb]\label{Figure1}
\includegraphics[scale=.75]{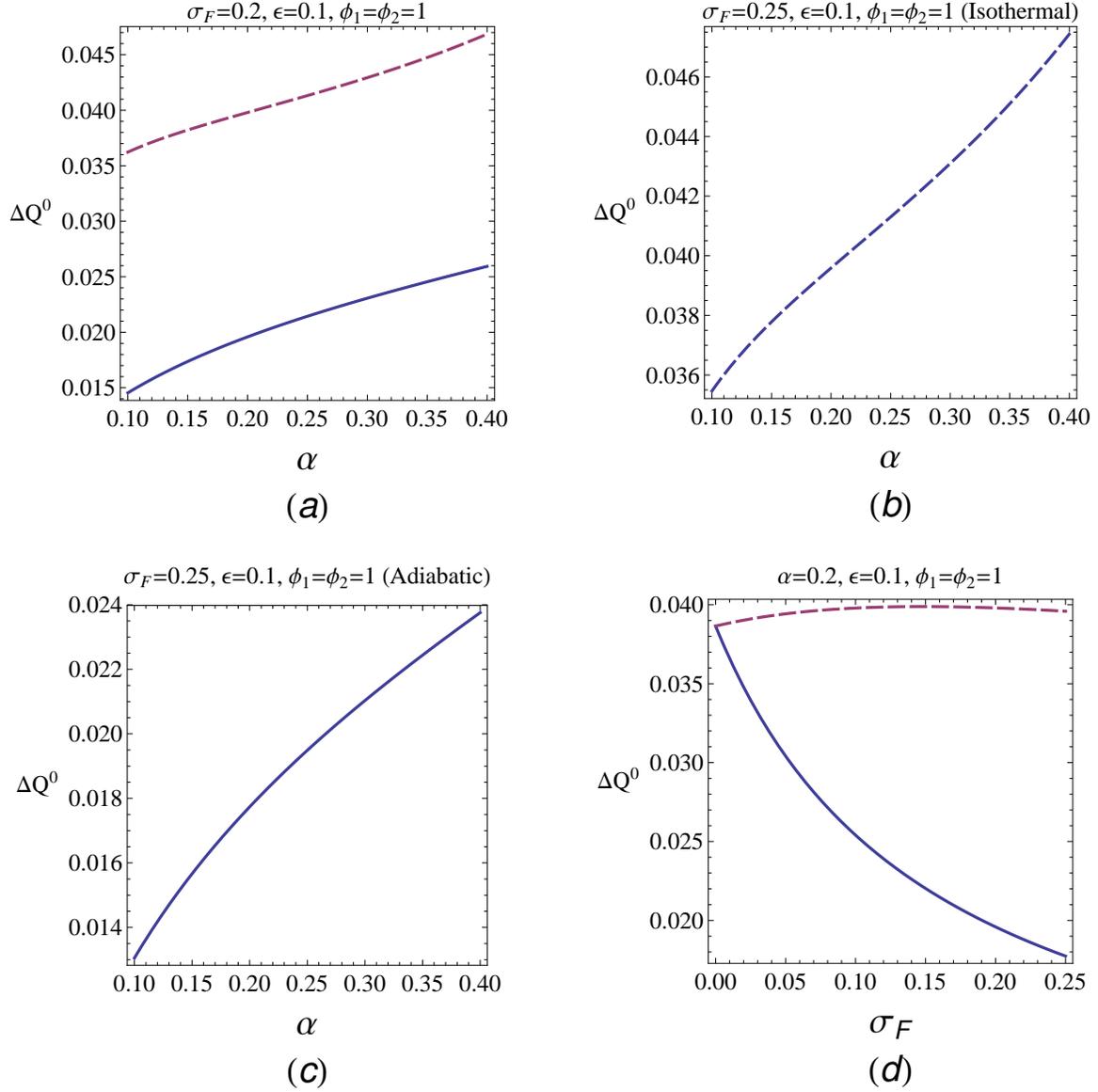}\caption{The variations of collision phase-shift with respect to the fractional positron to electron density, $\alpha$ (Figs. 1(a-c)), and the ion-temperature to electron Fermi-temperature ratio, $\sigma_F$ (Fig. 1(d)), for other fixed plasma parameters. The solid curves correspond to the adiabatic cases and the dashed curves refer to the isothermal ones.}
\end{figure}

\newpage


\begin{thebibliography}{9}

\bibitem {Chandra} S. Chandrasekhar, "An Introduction to the Study of Stellar Structure", Chicago, Ill., (The University of Chicago press), (1939) Ch. 4.
\bibitem {Manfredi} G. Manfredi, Fields Inst. Commun. \textbf{46} 263(2005).
\bibitem {Gardner} C. Gardner, SIAM, J. Appl. Math. \textbf{54} 409(1994).
\bibitem {Shukla} Shukla P K and Eliasson B, "Nonlinear aspects of quantum plasma physics" Phys. Usp. \textbf{53} 51(2010).
\bibitem {landau} L. D. Landau and E. M. Lifshitz, "Statistical Physics", Part I, Pergamon, Oxford, (1978).
\bibitem {Akbari} M. Akbari-Moghanjoughi, Phys. Plasmas \textbf{17}, 114701(2010).
\bibitem {Yan} Yan-Xia Xu, Zong-Ming Liu, Mai-Mai Lin, Yu-Ren Shi, Jian-Min Chen and Wen-Shan Duan, Phys. Plasmas \textbf{18}, 052301(2011).
\bibitem {Elshamy} E. F. El-Shamy, W. M. Moslem and P. K. Shukla, Phys. Lett. A, \textbf{374} (2009)290–293.
\bibitem {Khan1} S. A. Khan and Q. Haque, Chin. Phys. Lett., \textbf{25} 4329(2008).
\bibitem {Khan2} S. A. Khan, S. Mahmood, Arshad M. Mirza, Chin. Phys. Lett., \textbf{25} 045203(2009).
\bibitem {Ali} S. Ali, W. M. Moslem, P. K. Shukla, and R. Schlickeiserd, Phys. Plasmas \textbf{14} 082307(2007).
\bibitem {Moslem} W. M. Moslem, R. Sabry, and P. K. Shukla, Phys. Plasmas \textbf{17} 032305(2010).
\bibitem {Mushtaq1} A. Mushtaq and S. A. Khan, Phys. Plasmas \textbf{14} 052307(2007).
\bibitem {Mushtaq2} S. A. Khan and W. Masood, Phys. Plasmas \textbf{15} 062301(2008).
\bibitem {Masood} W. Masood, Arshad M. Mirza and M. Hanif, Phys. Plasmas \textbf{15} 072106(2008).
\bibitem {Abdelsalam} U. M. Abdelsalam, W. M. Moslem, P. K. Shukla, Phys. Lett. A, \textbf{372} 4057(2008).
\bibitem {Yue} Yue-yue Wang, Jie-fang Zhang, Phys. Lett. A, \textbf{372} 3707(2008).
\bibitem {Saeed} Saeed-ur-Rehman, N. Akhtar, and Asif Shah, Phys. Plasmas \textbf{18} 032303(2011).
\bibitem {Zhenya} Zhenya Yan, Phys. Lett. A, \textbf{373} 2432(2009).
\bibitem {Akhtar} N. Akhtar and S. Hussain, Phys. Plasmas \textbf{18} 072103(2011).

\end{thebibliography}
\end{document}